\documentclass[10pt,prx,aps,twocolumn,superscriptaddress,floatfix]{revtex4-1}

\pdfoutput=1

\usepackage{graphicx}
\usepackage{amssymb}
\usepackage{bm}
\usepackage{color}

\begin{document}

\title{Dynamic structure factor from real time evolution and exact correction
  vectors with matrix product states}

\author{Ling Wang}
\email{lingwang@csrc.ac.cn}
\affiliation{Beijing Computational Science Research Center, 10 East Xibeiwang Rd, Beijing 100193, China}

\author{Hai-Qing Lin}
\email{haiqing0@csrc.ac.cn}
\affiliation{Beijing Computational Science Research Center, 10 East Xibeiwang Rd, Beijing 100193, China}

\date{\today}

\begin{abstract}
  We propose two complimentary numerical methods for rigorous computation of
  dynamic structure factor at zero temperature. One solution is to solve
  Schrodinger equation using time dependent variational principle (TDVP) with
  matrix product states (MPSs) proposed by Haegeman el al. [Physical Review B
  94, 165116], and Fourier transform correlation functions in time to
  frequency domain. Another way is to directly compute the transition rate
  between ground state and several low lying momentum eigenstates that are
  accessed using MPSs method amended with momentum filtering process. We
  benchmark both methods on a spin-$1/2$ Antiferromagnetic (AF) Heisenberg
  chain with periodic boundary condition. With finite size scaling analysis,
  the asymptotic line shape as a function of $\omega$ can be reproduced at
  various momentum $k$ with both methods.
\end{abstract}
 
\maketitle

\section{\label{intro}Introduction}
Dynamic structure factor is an important physical quantity observed directly
in inelastic neutron scattering and resonating inelastic X-ray
spectroscopy~\cite{vanhove.pr95.249}. Numerical exact computation of them is
extremely important for experimental data analysis and theoretical
predictions. The definition of dynamic structure factor involves either real
time evolution operators $e^{\pm iHt}$, which leads to a sign problem, or a
set of exact eigenstates that is often inaccessible for non-integrable
systems. Recent numerical advances proposed several remedies to the above
problem. For a sign free Hamiltonian, analytical continuation with constraint
can generate reliable spectral functions with certain known physical
input~\cite{shao.prx7.041072}. On the other hand, for frustrated systems,
approximated methods have to be used. Variational methods, such as
Variational Monte Carlo (VMC)
method~\cite{li.prb81.214509,ferrari.prb97.235103,ferrari.prb88.100405},
matrix product states (MPSs)
method~\cite{Haegeman.prb85.100408,Vanderstraeten.prb92.125136,Vanderstraeten.prb93.235108,Zauner-stauber.prb97.045145,Zauner-stauber.prb97.235155},
and projected entangled pair states (PEPSs)
method~\cite{vanderstraeten.arxiv1809.06747}, have made translational
invariant variational ansatz for single or two particle excited states, from
which the transition rate between ground state and excited states can be
calculated. These methods are strongly restricted by their number of
variational parameters, yet they are efficient in getting all momentum
eigenstates and eigen energies within a single run. Apart from variational
approaches, exact diagonalization within certain quantum
sectors~\cite{Luscher.prb79.195102} is still a simple and practical choice to
get important physical insight, but it can not handle relatively large system
sizes. Follow the streamline of working in a Krylov space, Gagliano et
al. proposed continued fraction method~\cite{Gagliano.PRL59.2999}. With the
success of Density Matrix Renormalization Group (DMRG) method, it was
naturally adapted within a DMRG approach~\cite{Hallberg.prb52.r9827,
  White.PRB60.335}, where the exponential wall effect can be
significant. Subsequently, correction vectors were introduced as a remedy,
however with the expenses that each $\omega$ has to be targeted
separately~\cite{White.PRB60.335,Jeckelmann.PRB66.045114,Nocera.PRE94.053308}. To
circumvent this issue, Chebyshev MPSs (ChMPS) approach was proposed
~\cite{Holzner.PRB83.195115,Bruognolo.PRB94.085136,Xie.PRB97.075111} to
compute the full spectral functions. However the intrinsic limitation of ChMPS method
is at the size of the subspace expand by a set of orthogonal finite bond
dimensional MPSs~\cite{Haegeman.PRB88.075133}.

Another approach within the MPSs/DMRG framework is to compute the real time
evolution of an initial state using so called time dependent DMRG (tDMRG)
method~\cite{White.PRB48.3844,White.PRL93.076401,Pereira.PRL100.027206,White.PRB77.134437,VidalTEBD,Zaletel.PRB91.165112}. The
accuracy of this method depends on evolution time step $\tau$, the order of
Trotter-Suzuki expansion, and the number of Schmidt states $m$ kept. Usually
with $\tau$ ranges from $0.01$ to $0.1$ and the expansion order from second
to fifth, a maximum evolution time of $20$ to $40$ (in unit of coupling
strength) can be reached up-to $m=1200$~\cite{White.PRB77.134437}. One
obstacle of tDMRG method is the balancing between the order of expansion to
operator $e^{-i\tau H}$ expressed by a Matrix Product Operator (MPO) and the
efficiency (bond dimension) of
it~\cite{White.PRB77.134437,Zaletel.PRB91.165112}. A recent development in
the MPSs context for real time evolution~\cite{tdvpref} overcomes the above
problem and brings additional benefits. It deals with real time evolution by
numerically computing exponential of Hamiltonian (without Trotter-Suzuki
expansion), whose procedure is similar to the ground state search DMRG
algorithm, therefore the computational cost is also comparable to that of the
standard algorithm. The second advantage is that energy conservation for
unitary transformation is preserved, in contrary to the tDMRG method. Most
importantly, the optimization at each step $\tau$ explores the entire MPSs
manifold, which can be dynamically expanded when needed. This can be thought
of as the wavefunction $|\psi(t)\rangle$ is fully optimized in a subspace
spanned by all possible orthogonal states of MPSs with bond dimension
$m$. The detail of this method is re-visited in Sec.~\ref{tdvpmethods}. We
propose using this method for computing correlation functions in time, then
Fourier transform into frequency domain to get the spectral functions.

Come back to correction vectors in DMRG algorithm for spectral functions. If
a correction vector is an eigenstate, thus calculated spectral function is
exact. We here propose a new generic MPSs/DMRG algorithm to numerically
exactly compute physically important correction vectors. This benefits from
the MPSs formalism, which can deal with non-local operations, such as global
projection operator or translation operator, and can efficiently express them
as local matrix product operators (MPO). These non-local operations if
included as a constraint in the wavefunction, or added as extra terms in the
Hamiltonian, can lead to energy level reshuffle~\cite{lingPRL2018} or
reshaping. Inspired by the excited state MPSs/DMRG algorithm proposed by Wang
and Sandvik~\cite{lingPRL2018}, a MPSs/DMRG algorithm amended with momentum
filtering process is proposed to explicitly compute several energy
eigenstates with a wave momentum $\mathbf{k}$. This leads to direct
computation of the spectral weight (transition rate) of low-lying
excitations, for example the des Cloizeaux-Pearson (dCP)
states~\cite{Cloizeaux.PR128.2131} in the spin-$1/2$ Antiferromagnetic
Heisenberg chain. The rest of the paper is organized as following: after
revisit the TDVP in MPSs manifold in Sec.~\ref{tdvpmethods}, we discuss the
Fourier transformation needed to obtain spectral functions in
Sec.~\ref{ftrans}, where an alternative way to compute the spectral weight,
is also discussed. Sec.~\ref{benchmark} is devoted for a benchmark
demonstration of both methods. Conclusions and remarks are given in
Sec.~\ref{conclusion}.

\begin{figure}
\begin{center}
\includegraphics[width=\columnwidth]{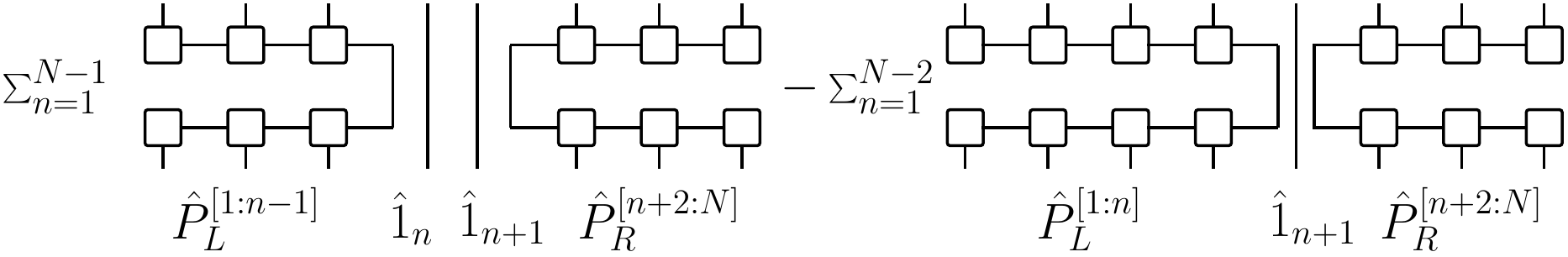}
\vskip-1mm
\caption{Trotter decomposed tangent space projector
  $\mathcal{P}_{\mathcal{T}}$ in a MPS manifold.}
\label{fig1}
\end{center}
\vskip-2mm
\end{figure}

\section{\label{tdvpmethods}Time dependent variational principle applied to
  matrix product states manifold}
Time dependent variational principle was formulated recently to solve the
Schrodinger equation approximately in a variational manifold $\mathcal{M}$ of
MPSs~\cite{tdvpref}. Consider Schrodinger equation, taking $\hbar=1$, 
\begin{equation}
\label{scheq0}
\frac{d\psi(t)}{dt}=-iH\psi(t).
\end{equation}
When restricted to a variation manifold $\mathcal{M}$, it becomes
\begin{equation}
\label{scheq}
\frac{du(t)}{dt}=-i\mathcal{P}_{\mathcal{T}_u}Hu(t),
\end{equation}
where $u(t)$ is an approximate solution to $\psi(t)$ within the manifold
$\mathcal{M}$, and $\mathcal{T}_u$ denotes a tangent space at
$u(t)$. Projection to the tangent space $\mathcal{P}_{\mathcal{T}_u}$
guarantees that, under unitary evolution, the norm of wavefunction doesn't
change. Ref.~\cite{tdvpref} has shown how to Trotter decompose a tangent
space projector rather than the Hamiltonian terms. Here we briefly outline
this idea following Ref.~\cite{tdvpref}. The wavefunction at time $t$ can be
written as
\begin{equation}
\label{psimps}
|\psi[A]\rangle=\sum_{\{s=1\}}^d \text{Tr}[A^{s_1}_1A^{s_2}_2\cdots
A^{s_N}_N]|s_1s_2\cdots s_N\rangle,
\end{equation}
which depends on a set of site-dependent matrices $A^{s_i}_i$. It is
convenient to rewrite Eq.~\ref{psimps} in a mixed canonical form
\begin{eqnarray}
\nonumber
|\psi[A]\rangle=&\sum_{\alpha\beta
                  s_ns_{n+1}}[A^C_nA^C_{n+1}]^{s_ns_{n+1}}_{\alpha\beta}\times\\
&|\Phi_{L\alpha}^{[1:n-1]}\rangle|s_n\rangle |s_{n+1}\rangle|\Phi_{R\beta}^{[n+2:N]}\rangle,\hfill
\end{eqnarray}
where $|\Phi_{L\alpha}^{[1:n-1]}\rangle$ ($|\Phi_{R\beta}^{[n+2:N]}\rangle$)
is a set of orthonormal basis for the left (right) block of the lattice
respect to the center two sites $n$ and ${n+1}$, $A^C_n$ and $A^C_{n+1}$ are
the center matrices of the two-site mixed canonical form. The tangent space
of above wavefunction is defined as
\begin{eqnarray}
\nonumber
|\Theta[B]\rangle=&\sum_{n=1}^{N-2}\sum_{\alpha\beta
  s_ns_{n+1}}[B_nB_{n+1}]^{s_ns_{n+1}}_{\alpha\beta}\times\\
&|\Phi_{L\alpha}^{[1:n-1]}\rangle|s_n\rangle|s_{n+1}\rangle|\Phi_{R\beta}^{[n+2:N]}\rangle.
\end{eqnarray}
Under the "left gauge fixing condition"
\begin{equation}
\label{constraint}
\sum_{s_n\beta}[A_L]_{\beta\alpha}^{s_n}(n)B_{\beta\alpha^{\prime}}^{s_n}(n)=0,\quad \forall n=1,\dots,N-1,
\end{equation}
where $A_L(n)$ denotes the left canonical form of $A_n$, one can check that
$\langle \psi[A]|\Theta[B]\rangle=0$. To minimize the distance of an
arbitrary vector $|\Xi\rangle$ with tangent vector $|\Theta[B]\rangle$ under
constraint Eq.~\ref{constraint}, the projector is derived as
\begin{eqnarray}
\nonumber
\mathcal{P}_{\mathcal{T}}=&\sum_{n=1}^{N-1}\hat{P}_L^{[1:n-1]}\otimes\hat{1}_n\otimes\hat{1}_{n+1}\otimes\hat{P}_R^{[n+2:N]}\\
\label{projector}
&-\sum_{n=1}^{N-2}\hat{P}_L^{[1:n]}\otimes\hat{1}_{n+1}\otimes\hat{P}_R^{[n+2:N]}\hfill,
\end{eqnarray}
where $\hat{P}_L^{[1:n-1]}$ and $\hat{P}_R^{[n+2:N]}$ are defined as in
Fig.~\ref{fig1}. Once Trotter decompose the tangent space projector into
parts, one can then integrate one by one following naturally the sweeping
order of the standard DMRG algorithm.

The two-site algorithm can be organized as following: (1) prepare initial
state in a right canonical form. (2) For any $n=1,\cdots,N-2$ integrate the
Schrodinger equation in a subspace centered around $n$ and ${n+1}$
\begin{equation}
\frac{dA_C(n,t)}{dt}=-iH_{\text{eff}}A_C(n,t),
\end{equation}
where $A_C(n,t)=\sum_{\alpha\beta s_ns_{n+1}}[A^C_n(t)A^C_{n+1}(t)]^{s_ns_{n+1}}_{\alpha\beta}$. The solution is
\begin{equation}
A_C(n,t+\tau)=e^{-iH_{\text{eff}}\tau}A_C(n,t).
\end{equation}
(3) Reformulate the wavefunction as a one-site ($A^C_{n+1}$) centered mixed
canonical form to integrate the Schrodinger equation backward in time
\begin{equation}
\frac{dA^C_{n+1}(t+\tau)}{dt}=-iH_{\text{eff}}A^C_{n+1}(t+\tau),
\end{equation}
where the solution is
\begin{equation}
A^C_{n+1}(t)=e^{-iH_{\text{eff}}(-\tau)}A^C_{n+1}(t+\tau).
\end{equation}
Note that step (2) and (3) together complete evolving $A_L(n,t)$ to
$A_L(n,t+\tau)$, with $A^C_{n+1}(t)$ yet to be updated by moving the center one
site to the right. (4) Finish right-ward sweep by executing step (2) for
$n=N-1$. Steps (1) to (4) all together evolve $|\psi[A](t)\rangle$ to
$|\psi[A](t+\tau)\rangle$. Similarly, one can perform a left-ward sweep.  A
full (left-right then right-left) DMRG sweep evolves $|\psi[A](t)\rangle$
forward in time by $2\tau$.

\begin{figure}
\begin{center}
\includegraphics[width=\columnwidth]{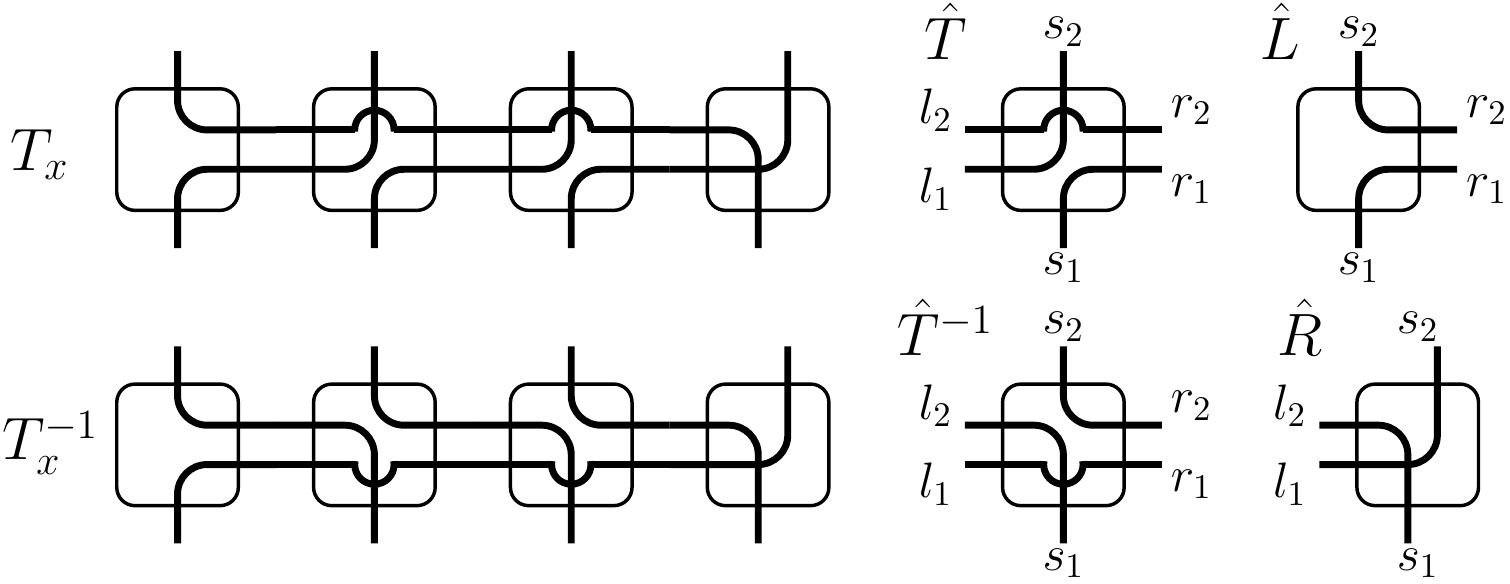}
\vskip-1mm
\caption{Demonstration of translation operators $T_x$ and $T_x^{-1}$ in a
  form of matrix product operator.}
\label{mpotx}
\end{center}
\vskip-2mm
\end{figure}

\section{\label{ftrans}Dynamic structure factor and transition rate matrix}
Dynamic structure factor can be expressed as Fourier transformation of space
and time separated correlation function
\begin{equation}
S^{\alpha\beta}(\mathbf{k},\omega)=\frac{1}{N}\sum_{\mathbf{r},\mathbf{l}}e^{-i\mathbf{k}\cdot(\mathbf{r}-{\mathbf{l}})}\int_{-\infty}^{\infty}dte^{i\omega t}\langle 0|S^{\alpha}_{\mathbf{r}}(t)S^{\beta}_{\mathbf{l}}(0)|0\rangle,
\end{equation}
where $|0\rangle$ denotes the ground state, $\alpha,\beta=x,y$ or $z$, and
$\mathbf{r}$ is a coordinate index runs over all sites in the system. The
space Fourier transformed operator
$S_{\mathbf{k}}^{\alpha}=\frac{1}{\sqrt{N}}\sum_{\mathbf{r}}e^{i\mathbf{k}\cdot\mathbf{r}}S^{\alpha}_{\mathbf{r}}$
can be conveniently expressed as a matrix product operator (MPO) of bond
dimension 2, where the left and right boundary MPO can be written as
$(e^{i\mathbf{k}\cdot\mathbf{r}_1}\hat{S}_{\mathbf{r}_1}^{\alpha},\hat{1})$ and
$(\hat{1},e^{i\mathbf{k}\cdot\mathbf{r}_N}\hat{S}_{\mathbf{r}_N}^{\alpha})^T$
respectively, and the bulk MPO is written as
\begin{eqnarray}
\left(
\begin{array}{cc}
\hat{1}&0\\
e^{i\mathbf{k}\cdot\mathbf{r}_i}
\hat{S}_{\mathbf{r}_i}^{\alpha}&\hat{1}
\end{array}
\right).
\end{eqnarray}
The dynamic structure can be re-expressed as
\begin{eqnarray}
\label{strucfac1}
\nonumber
S^{\alpha\beta}(\mathbf{k},\omega)&=&\int_{-\infty}^{\infty}dte^{i\omega
  t}\langle 0|S_{\mathbf{-k}}^{\alpha}(t)S_{\mathbf{k}}^{\beta}|0\rangle\\
\label{strucfac2}
&=&\int_{-\infty}^{\infty}dte^{i(\omega+E_0)t}\langle 0|S_{-\mathbf{k}}^{\alpha}e^{-iHt}S_{\mathbf{k}}^{\beta}|0\rangle.
\end{eqnarray}

Insert a complete set of basis $\sum_i|i\rangle\langle i|$ into
Eq.~\ref{strucfac2}, the dynamic structure factor can be written as following
\begin{equation}
S^{\alpha\beta}(\mathbf{k},\omega)=2\pi\sum_i\langle
0|S_{\mathbf{-k}}^{\alpha}|i\rangle\langle i|S_{\mathbf{k}}^{\beta}|0\rangle \delta(\omega-\omega_i),
\end{equation}
where $\omega_i=E_i-E_0$, $E_i=\langle i|H|i\rangle$ and
$E_0=\langle 0|H|0\rangle$. The dynamic structure factor reduces to a set of
poles (delta function) with transition rate
$M_i^{\alpha\beta}(\mathbf{k},\omega)\equiv \langle
0|S_{\mathbf{-k}}^{\alpha}|i\rangle\langle
i|S_{\mathbf{k}}^{\beta}|0\rangle$, which is nonzero only when eigen state
$|i\rangle$ has momentum $\mathbf{k}$.

In the MPSs formalism, momentum eigenstates can be selected by adding an
extra cost $H_{\lambda}$ in the Hamiltonian
\begin{eqnarray}
\nonumber
H_{\lambda}&=&\lambda \left[\Big(\frac{T_{\alpha}+T^{-1}_{\alpha}}{2}-\text{cos}k_{\alpha}\Big)^2+\Big(\frac{T_{\alpha}-T_{\alpha}^{-1}}{2i}-\text{sin}k_{\alpha}\Big)^2\right]\\
&=&-\lambda\Big( e^{ik_{\alpha}}T_{\alpha} +e^{-ik_{\alpha}}T_{\alpha}^{-1}\Big)+\text{const.},
\end{eqnarray}
where $T_{\alpha}$ is a translation operator in $\alpha=x$ or $y$ direction,
$\lambda$ is a large number to favor energy eigenstate with wave momentum
$k_{\alpha}$. The translation operator can be written efficiently as a MPO of
bond dimension $d^2$, as illustrated in Fig.~\ref{mpotx}, with
$\hat{L}=\hat{1}_{s_1r_1}\otimes \hat{1}_{s_2r_2}$,
$\hat{R}=\hat{1}_{s_1l_2}\otimes \hat{1}_{s_2l_1}$,
$\hat{T}=\hat{1}_{s_1r_1}\otimes \hat{1}_{s_2l_1}\otimes \hat{1}_{l_2r_2}$
and
$\hat{T}^{-1}=\hat{1}_{s_1l_2}\otimes \hat{1}_{s_2r_2}\otimes
\hat{1}_{l_1r_1}$, where $\hat{1}$ is a $d\times d$ identity matrix.

In the selected momentum $k_{\alpha}$ sector, the ground state as well as
several low lying excited states can be computed successively using
Hamiltonian
\begin{equation}
H_{j}^{\prime}=H+H_{\lambda}-\sum_{i=0}^{j-1}E_{i}|i\rangle\langle i|\quad (i<j),
\end{equation}
where $|i\rangle$ (for any $i<j$, $E_i<E_j$) are energy eigenstates with wave
momentum $k_{\alpha}$ that are pre-computed before targeting the next state
$|j\rangle$. The transition rate
$M_i^{\alpha\beta}(\mathbf{k},\omega)\equiv \langle
0|S_{\mathbf{-k}}^{\alpha}|i\rangle\langle i|S_{\mathbf{k}}^{\beta}|0\rangle$
therefore can be computed directly using eigenstates $|0\rangle$ and
$|i\rangle$.

\section{\label{benchmark}Dynamic structure factor of spin-$1/2$
  Antiferromagnetic Heisenberg chain}
\subsection{Real time evolution}
\begin{figure}
\begin{center}
\includegraphics[width=6.5cm]{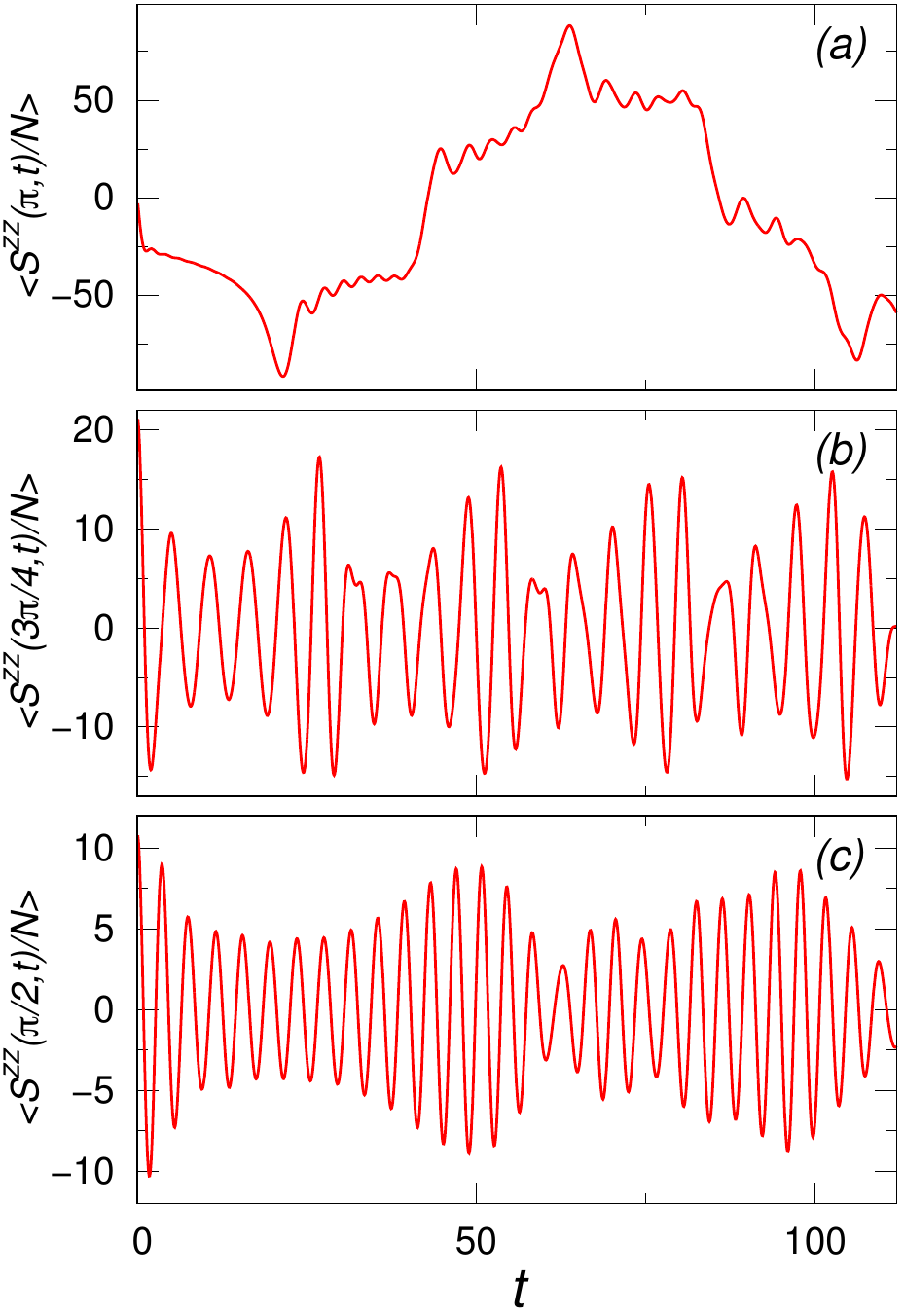}
\vskip-1mm
\caption{Real part of the spin-spin correlation functions in time at momentum
$k=\pi,\frac{3\pi}{4}$, and $\frac{\pi}{2}$ for size $N=64$ chain.}
\label{tevol}
\end{center}
\vskip-2mm
\end{figure}

We study the spin-$1/2$ Antiferromagnetic Heisenberg chain of $N$ sites with
periodic boundary condition
\begin{equation}
H=\sum_{i=1}^N\mathbf{S}_i\cdot\mathbf{S}_{i+1}.
\end{equation}
The dynamic structure factor of this model has been well studied with Bethe
Ansatz~\cite{Yamada.PTPJ41.880,Cloizeaux.PR128.2131}, symmetries and quantum groups
analysis~\cite{Muller.PRB24.1429,Bougourzi.PRB54.r12669,Karbach.PRB55.12510},
and various numerical methods, such as time dependent DMRG
method~\cite{Pereira.PRL100.027206}, the Krylov-space approach with
correction vectors~\cite{Nocera.PRE94.053308}, the MPS-based Chebyshev
expansion method~\cite{Xie.PRB97.075111}. We study this model by computing
the spin correlation function in real time, {\it i.e.}  the integrand
$\langle 0|S_{\mathbf{-k}}^{\alpha}(t)S_{\mathbf{k}}^{\beta}|0\rangle$ of
Eq.~\ref{strucfac1}, then Fourier transform into frequency domain to obtain
the dynamic structure factor $S(\mathbf{k},\omega)$. Since the Hamiltonian is
one dimensional, we here and after replace the wave momentum $\mathbf{k}$ by
a scalar $k$ for simplicity. The integrand is calculated by separately
evolving $|S_k^{\beta}(t/2)\rangle=e^{-i\beta Ht/2}S_k^{\beta}|0\rangle$ and
$|S_k^{\alpha\dagger}(-t/2)\rangle=e^{i\beta
  Ht/2}S_k^{\alpha\dagger}|0\rangle$ in time, and taking the inner product of
them to get the time dependent correlation function
\begin{eqnarray}
\nonumber
\langle S^{\alpha\beta}(k,t)\rangle&=&\langle 0|S_{-k}^{\alpha}(t)S_{k}^{\beta}|0\rangle\\
\label{strucfac3}
&=&e^{iE_0t}\langle S_{k}^{\alpha\dagger}(-t/2)|S_k^{\beta}(t/2)\rangle.
\end{eqnarray}
The correlation functions $\langle S^{zz}(k,t)\rangle$ for a size $N=64$
chain at momentum $k=\pi,\frac{3\pi}{4}, \frac{\pi}{2}$ are shown in
Fig.~\ref{tevol}. Here we take $\tau=0.02$ and iterate $1400$ times to reach
$T_{\text{max}}=112$. The bond dimension is dynamically adjusted such that the
error throw away in a single SVD $\epsilon<10^{-7}$. For $N=64$, maximum bond
dimension can reach $m=2000$.

\begin{figure}
\begin{center}
\includegraphics[width=6.5cm]{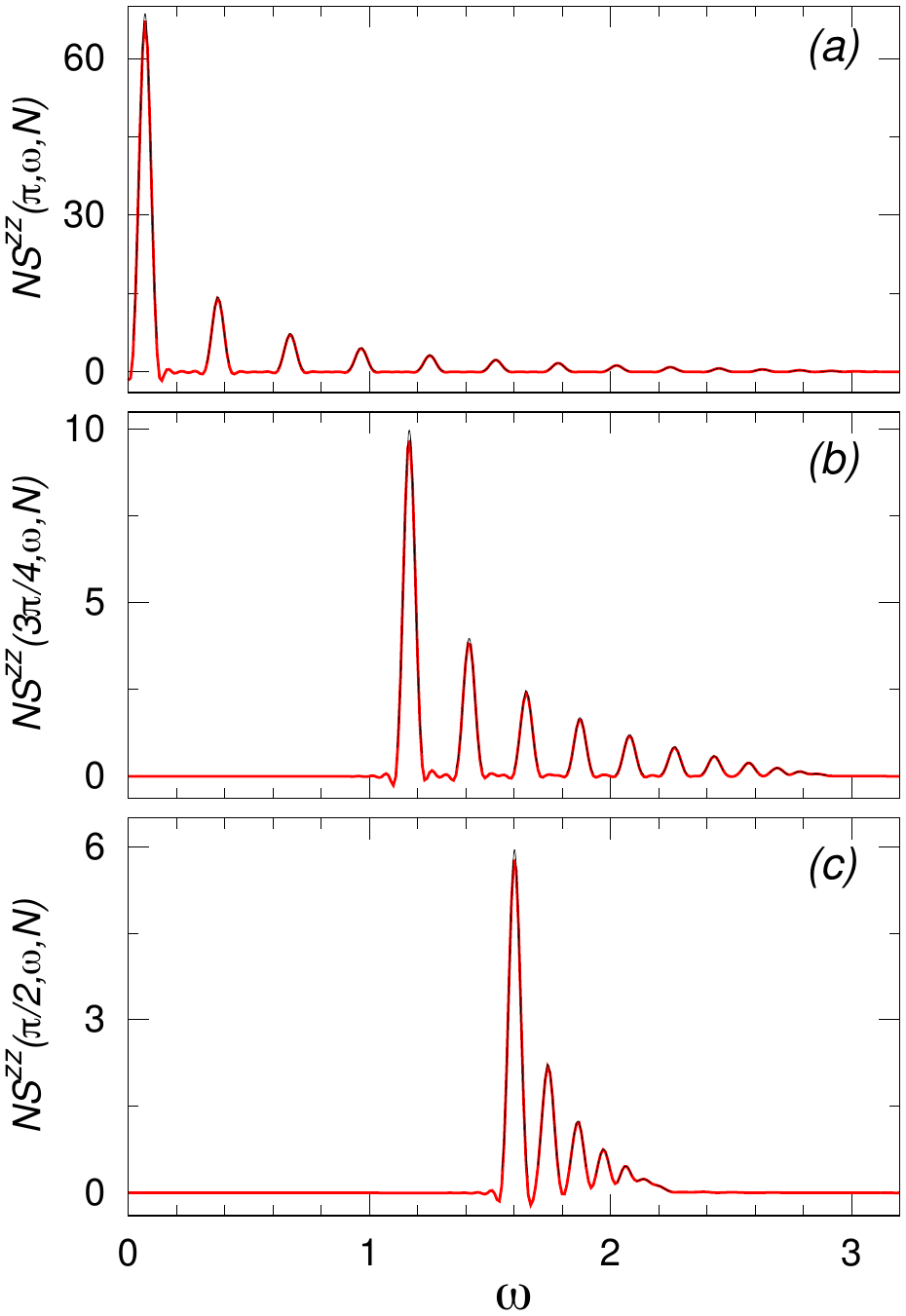}
\vskip-1mm
\caption{Size scaled dynamic structure factor $S^{zz}(k,\omega,N)$ obtained
  from Fourier transformation spin correlation functions in time shown in
  Fig.~\ref{tevol} for $k=\pi,\frac{3\pi}{4}$, and $\frac{\pi}{2}$. Black
  solid lines are Gaussian function fits of the form
  $a_i\text{exp}{(-\frac{(\omega-\omega_i)}{2b_i^2})}$ for each peak at
  $\omega_i$.}
\label{spectraplot}
\end{center}
\vskip-2mm
\end{figure}

When Fourier transform time dependent correlation functions into frequency
domain, we multiply the integrand by a contour
$1+\text{cos}\big(\frac{\pi t}{T_{\text{max}}}\big)$, therefore, each delta
function peak becomes a Gaussian with a broadening
$\approx \frac{1}{T_{\text{max}}}$. Fig.~\ref{spectraplot} illustrates the
dynamic structure factors of $N=64$ chain at momentum
$\pi,\frac{3\pi}{4},\frac{\pi}{2}$ obtained from the Fourier transform of
spin correlation functions in Fig.~\ref{tevol}. We fit each peak in the
spectral function with a Gaussian $a_ie^{-\frac{(\omega-\omega_i)}{2b_i^2}}$
(shown in black solid lines in Fig.~\ref{spectraplot}), where $b_i\approx 0.02$
for all fits.

\subsection{Comparison to two-spinon analytical transition rates}
For the AF Heisenberg chain, its lowest excited states are the famous des
Cloizeaux-Pearson (dCP) triplets~\cite{Cloizeaux.PR128.2131} 
\begin{equation}
\omega_L(k)=\frac{\pi}{2}|\text{sin}k|.
\end{equation}
Bethe Ansatz approach revealed an extended two-spinon
continuum whose lower boundary is the dCP expression
and upper boundary is given by~\cite{Yamada.PTPJ41.880}
\begin{equation}
\omega_U(k)=\pi|\text{sin}k/2|.
\end{equation}
By approaches based on concept of infinite dimensional symmetries developed
in the context of quantum groups, exact two-spinon dynamic structure factor
has been
derived~\cite{Muller.PRB24.1429,Bougourzi.PRB54.r12669,Karbach.PRB55.12510},
the asymptotic behavior at $k=\pi$ and $\omega\to 0$ is
\begin{equation}
S^{\alpha\beta}(\pi,\omega)\propto
  \frac{1}{\omega}\sqrt{\text{ln}\frac{1}{\omega}},
\end{equation}
whereas for all other $k$ at $\omega\to\omega_L$ is
\begin{equation}
S^{\alpha\beta}(k,\omega)\propto  \frac{1}{\sqrt{\omega-\omega_L(k)}}\sqrt{\text{ln}\frac{1}{\omega-\omega_L(k)}}.
\end{equation}
To verify the above asymptotic behaviors using finite size dynamic structure
factor from real time evolution, one has to carefully normalize them. The sum
rule of the first moment, defined as
\begin{eqnarray}
\nonumber
K_1(k,N)&=&\int_0^{\infty}\frac{d\omega}{2\pi}\omega
            S^{\alpha\beta}(k,\omega)\\
&=&\frac{2E_0}{3N}(1-\text{cos}k),
\end{eqnarray}
where $E_0$ is the ground state energy, is known for all
$k$~\cite{Hohenberg.PRB10.128}. One can define a normalized dynamic structure
factor for given $k$ and $N$
\begin{eqnarray}
S^{\alpha\beta}(k,\omega,N)&=&\frac{K_1(k,N)}{\bar{K}_1(k,N)}\sum_ia_i(k,N)\delta(\omega-\omega_i),\\
\bar{K}_1(k,N)&\equiv&\sum_{i}\omega a_i(k,N)\delta(\omega-\omega_i),
\end{eqnarray}
where $a_i(k,N)$ is the height of fitted Gaussian function for each pole at
$\omega_i$ as in Fig.~\ref{spectraplot}.

\begin{figure}
\begin{center}
\includegraphics[width=6.5cm]{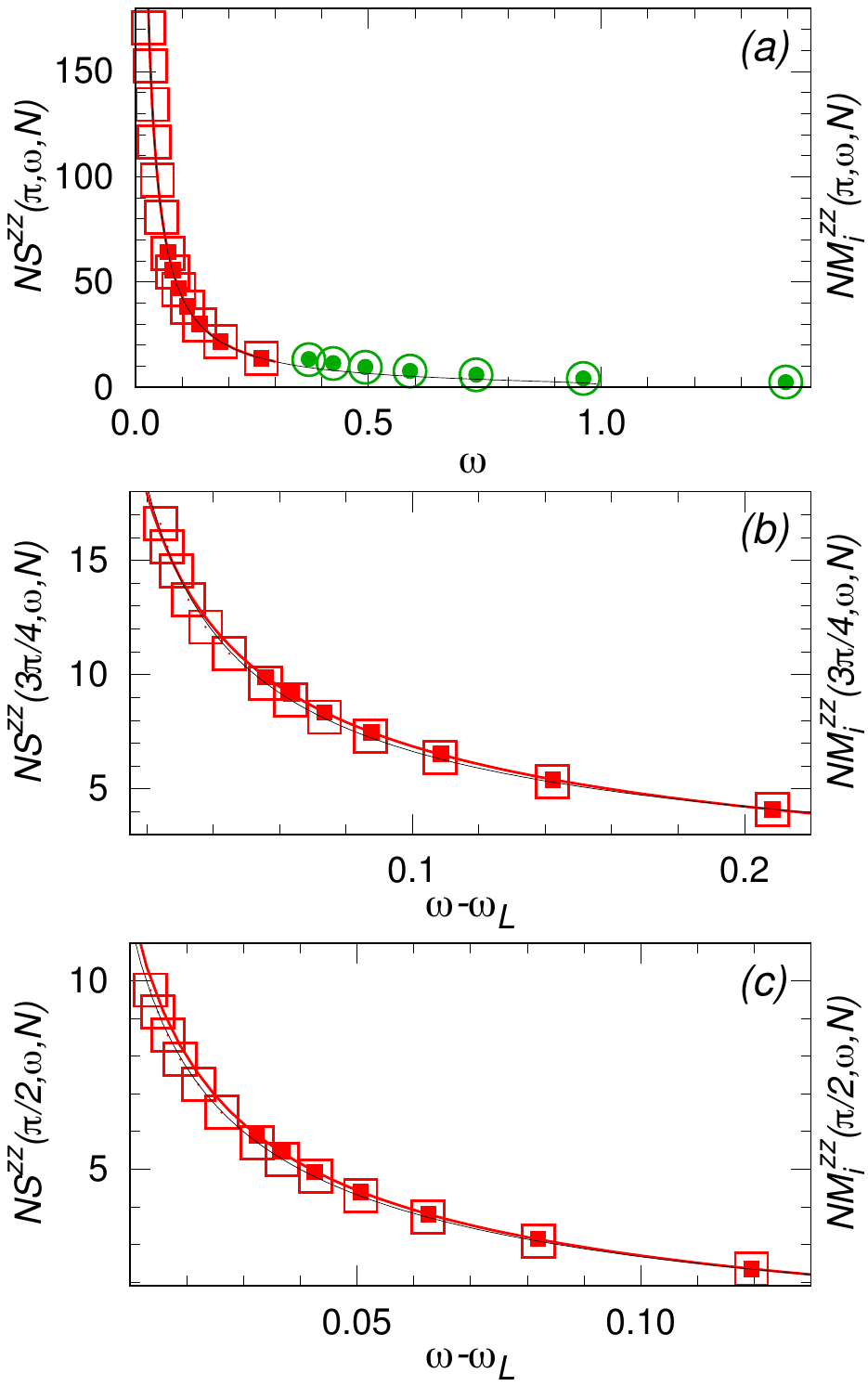}
\vskip-1mm
\caption{Size scaled and normalized dynamic structure factor from real time
  evolution in solid symbols, and size scaled transition rate from exact
  eigenstates in open symbols. Asymptotic behavior
  $\frac{1}{(\omega-\omega_L(k))^c}(\text{ln}\frac{1}{\omega-\omega_L(k)})^d$
  is fitted to above results, with $c,d$ the fitting parameters, red solid
  lines show fitting results from solid symbols, and black solid lines show
  fitting results from open symbols.}
\label{asympt}
\end{center}
\vskip-2mm
\end{figure}

The finite chain analysis~\cite{Karbach.PRB55.12510} showed that the scaled
transition rate $NM_i^{zz}(k,\omega,N)$ is a smooth function of $\omega$ by
varying size $N$. In Fig.~\ref{asympt} we plot the scaled and normalized
dynamic structure factor $NS^{zz}(k,\omega,N)$ for the first pole $\omega_1$
as a function of $\omega$ for sizes $N=64,56,48,40,32,24,16$ at momentum
$k=\pi,\frac{3}{4}\pi,\frac{1}{2}\pi$ in solid red
squares. Fig.~\ref{asympt}(a) also plots $NS^{zz}(k,\omega,N)$ for the second
pole $\omega_2$ at $k=\pi$ in solid green dots. Whereas the open symbols in
Fig.~\ref{asympt} are the scaled transition rate
$NM_i^{zz}(k,\omega,N)=N|\langle i|S_k^z|0\rangle|^2$ directly computed from
the lowest momentum eigenstates $|i\rangle$ for sizes
$N=160, 144, 128, 112, 96, 80, 64,56,48,40,32,24,16$ at
$k=\pi,\frac{3}{4}\pi,\frac{1}{2}\pi$ (in red open squares) and the second
lowest momentum eigenstates for sizes $N=64,56,48,40,32,24,16$ at $k=\pi$ (in
green open circles) via the momentum filtering method mentioned in
Sec.~\ref{ftrans}. Fig.~\ref{asympt}(a) indicates that the two complementary
methods produce exactly the same spectral functions, with solid and open
symbols lie on top of each other. Fitting with the asymptotic behavior
$\frac{1}{(\omega-\omega_L(k))^c}(\text{ln}\frac{1}{\omega-\omega_L(k)})^d$,
where $\omega_L(\pi)=0$, we found $c=1.00(3)$, $d=0.28(8)$, which is very
consistent with Bethe Ansatz analytical results
$c=1,d=0.5$. Fig.~\ref{asympt}(b-c) show that the weight of the first peak is
a bit over-counted in the Fourier transformation of correlation functions in
time for large system sizes. It means that longer evolution time and larger
bond dimension is needed to resolve small and close-by spectral peaks in
higher energy excitations. With the same analytical function we found in
Fig.~\ref{asympt}(b) for $k=3\pi/4$, $c=0.59(4),d=0.11(11)$ for open
symbols, $c=0.42(4),d=0.54(8)$ for solid symbols; while in
Fig.~\ref{asympt}(c) for $k=\pi/2$, $c=0.43(2),d=0.69(5)$ for open
symbols, and $c=0.40(7),d=0.83(17)$ for solid symbols.

\section{\label{conclusion}Remarks and discussions}
We proposed two different methods to compute dynamic structure factor, both
methods are formulated within the framework of matrix product states
(MPSs). One involves real time evolution using a recently proposal, where
time dependent variational principle (TDVP) is applied to
MPSs~\cite{tdvpref}. Another method directly target exact eigenstates in the
middle of the spectra by modifying Hamiltonian to favor eigenstates with wave
momentum $\mathbf{k}$. We applied both method to the spin-$1/2$
Antiferromagnetic Heisenberg chain. From real time evolution,
$T_{\text{max}}=112$ can be reached at a maximum bond dimension $m=2000$.
Linear prediction for real time correlation, which is believed to be
unreliable for complicated spectral functions, is not used here. Still the
Fourier transformation with a cut $T_{\text{max}}$ in time can rigorously
reproduce the exact spectral function for size $N=64$. Larger system sizes
can be studied with a slightly relaxed condition $\epsilon<10^{-6}$ for one
singular value decomposition (SVD). The computational cost of evolving a
wavefunction $2\tau$ forward in time is comparable to one full sweep in the
standard density matrix renormalization group (DMRG) algorithm. Given a
relatively small bond dimension $m=2000$, a much longer time $T_{\text{max}}$
can be reached compared to tDMRG algorithm formulated with matrix product
operator (MPO). Even though there is no explicit estimation of Trotter error
in orders of $\tau$ for Trotter expansion of the tangent space projector,
the forward plus backward evolution scheme make error in $\tau$ cancel thus
generate very accurate solution to the Schrodinger equation. On the other
hand, direct computation of energy eigenstates in the middle of the spectrum
proposed in this paper is way powerful than previous correction vector method
in DMRG, in a way that it can reshuffle and reshape the energy eigenvalues
such that around the target state, the density of states are much
smaller. With additional techniques, such as fixing total spin quantum number
with SU(2) symmetric MPSs/DMRG program, the convergence of eigenstates can be
even faster. The time evolution method, although powerful, can not evolve too
long with restricted bond dimension $m$. However it will predict a
qualitatively correct position $\omega_i$ of the important physics. Applying
shift-and-invert method using the proposed Hamiltonian in the main text can
precisely compute the corresponding eigenvector, and allows direct analysis
of other observable. The two methods combined together provide a numerical
powerful tool to explore excitations in quantum many-body systems.

Upon completing this manuscript, we saw manuscript~\cite{Paeckel}, which
shares similar idea with one of the two proposals within our manuscript.

\begin{acknowledgments} {\it Acknowledgments.---}We would like to thank
  F.~Verstraete, J.~Haegeman, R.~Mondaini, A.~Sandvik, H.~Shao, Y.-J.~Kao,
  Z.-X. Liu, and T.~Li for helpful discussions. L.W. is supported by the
  National Key Research and Development program of China (Grant
  No.~2016YFA0300600), the National Natural Science Foundation of China
  (Grant No.~NSFC-11874080 and No.~NSFC-11734002), the National Thousand
  Young Talents Program of China, and the NSAF Program of China (Grant
  No.~U1530401). H.Q.L. is supported by the National Natural Science
  Foundation of China (No.~NSFC-11734002) and the NSAF Program of China
  (Grant No.~U1530401). The calculations were partially carried out under a
  Tianhe-2JK computing award at the Beijing Computational Science Research
  Center (CSRC).
\end{acknowledgments}


\setcounter{page}{1}
\setcounter{equation}{0}
\setcounter{figure}{0}
\renewcommand{\theequation}{S\arabic{equation}}
\renewcommand{\thefigure}{S\arabic{figure}}

\end{document}